\documentclass[aps,prb,twocolumn,superscriptaddress,longbibliography]{revtex4-2}
\usepackage[colorlinks=true,linkcolor=blue,citecolor=blue]{hyperref}
\usepackage{amsmath,amssymb}
\usepackage[utf8]{inputenc}
\usepackage[T1]{fontenc}
\usepackage{url}
\usepackage{graphicx}
\usepackage{xcolor}
\usepackage{color}
\usepackage{ulem}
\usepackage{graphicx}

\begin{document}

\title{\texorpdfstring{Emergence of Non-Hermitian Magic Angles and Topological Phase Transitions in Twisted Bilayer $\alpha$-$T_3$ Lattices}{Emergence of Non-Hermitian Magic Angles and Topological Phase Transitions in Twisted Bilayer alpha-T3 Lattices}}

\author{Shaina Gandhi}
\email{shaina007@iitg.ac.in}
\affiliation{Department of Physics, Indian Institute of Technology Guwahati, Guwahati 781039, Assam, India}

\author{Gourab Paul}
\email{p.gourab@iitg.ac.in}
\affiliation{Department of Physics, Indian Institute of Technology Guwahati, Guwahati 781039, Assam, India}

\author{Srijata Lahiri}
\email{srijata.lahiri@iitg.ac.in}
\affiliation{Department of Physics, Indian Institute of Technology Guwahati, Guwahati 781039, Assam, India}

\author{Saurabh Basu}
\email{saurabh@iitg.ac.in}
\affiliation{Department of Physics, Indian Institute of Technology Guwahati, Guwahati 781039, Assam, India}

\begin{abstract}
 We investigate the flat-band properties and topological phase transitions in a non-Hermitian twisted bilayer $\alpha-T_3$ lattice. Here, non-Hermiticity is introduced via Hatano-Nelson-type asymmetric hopping, while an aligned hexagonal boron nitride substrate provides a staggered sublattice mass to the system. We find that the introduction of non-reciprocal hopping splits the conventional single magic angle into three distinct non-Hermitian magic angles (NHMAs). Unlike the exceptional magic angles driven by spectral singularities, these NHMAs host perfectly isolated flat bands where the real and imaginary parts of the bandwidth simultaneously vanish. By mapping the complex eigenspectrum across the moir\'e Brillouin zone, we show that the scattered energy eigenvalues coalesce into well-defined, closed loop-like structures as the non-Hermitian parameter strength increases, indicating emergence of a nontrivial point-gap topology and hence the non-Hermitian skin effect. Furthermore, we characterize the topological phases by computing the direct band gap and the biorthogonal Chern number. While the system exhibits a transition to a higher topological phase at weak non-Hermiticity, we demonstrate that stronger non-Hermiticity drives the gap-closing boundaries to merge and their topological charges to mutually annihilate. This convergence results in a trivial gap closing and a complete suppression of the intermediate topological phase, confirming that non-Hermiticity fundamentally plays a crucial role with regard to destabilizing the robust topological features of this moir\'e system.

\end{abstract}

\maketitle 
\section{Introduction}
\label{sec1}
Graphene, a one-atom-thick allotrope of carbon, has been extensively studied because its low-energy charge carriers behave as massless, chiral Dirac fermions. This linear energy dispersion qualifies them as pseudo-relativistic electrons \cite{RevModPhys.81.109}. However, overlaying two single layers with a relative twist angle creates a large-scale moir\'e superlattice, drastically modifying these electronic properties \cite{Li2010, PhysRevLett.99.256802, Bistritzer2011}. In the small twist angle regime, this moir\'e pattern causes a massive enlargement of the unit cell and induces band folding. As the twist angle approaches a specific magic angle, the interlayer coupling strongly suppresses the kinetic energy of the Dirac quasiparticles, causing their Fermi velocity to approach zero. This results in the emergence of nearly isolated, extremely flat bands that persist throughout the moir\'e Brillouin zone (mBZ) \cite{PhysRevB.82.121407}. In these flattened bands, electronic interactions become dominant, turning twisted bilayer graphene (tBG) into a fascinating platform for strongly correlated phenomena \cite{PhysRevB.102.045107, Cao2018, Choi2019}. This subtle interplay between band topology and moir\'e lattice symmetries gives rise to a rich landscape of exotic quantum phases such as topological Chern insulators \cite{Xie2021, Wu2021, Nuckolls2020}, ferromagnetic states \cite{Saito2021, Sharpe2019}, correlated nematics \cite{Cao2021_Nematic, RubioVerdu2022}, and superconductors that can be highly tuned by external parameters like doping and magnetic fields \cite{Cao2018_SC, Yankowitz2019, Saito2020, Oh2021, 1t3m-2sjn}. These twisted heterostructures offer immense tunability and can be realized across a broad spectrum of lattice types and stacking configurations \cite{Can2021, PhysRevB.104.L241107, Xiao2020, PhysRevB.109.155159, Chen2021, Park2021, Liu2020, PhysRevLett.122.086402, PhysRevB.111.045434, PhysRevB.111.045434}.

While twisting layers in moir\'e systems is a powerful way to create flat bands, similar flat bands can also emerge naturally in certain crystal geometries purely due to quantum interference \cite{PhysRevLett.62.1201, PhysRevLett.81.5888}. In these structures, the specific arrangement of atoms causes the electron wavefunctions to become localize certain lattice sites, which completely suppresses their kinetic energy. Classic examples of this phenomenon include the Kagome and Lieb lattices that host flat bands \cite{Kang2020,PhysRevB.99.125131, PhysRevB.111.115105, PhysRevLett.114.245504, Jiang2020}. Another highly tunable example is the $\alpha-T_3$ lattice, whose Hamiltonian emulates pseudospin-1 Dirac fermions \cite{PhysRevLett.112.026402, PhysRevB.92.245410}. Geometrically, this lattice consists of a standard graphene-like honeycomb structure (made of sublattices A and B) along with an extra atom (sublattice C) placed at the center of each hexagon. The parameter $\alpha$ acts as a tuning knob that smoothly transitions the system from regular graphene ($\alpha=0$) to the dice lattice limit ($\alpha=1$) by controlling the ease with which electrons can hop between the B and C atoms. In the dice limit, quantum interference completely disappears as the hopping amplitudes between the A-B and B-C sites become equal, trapping the states strictly at the central B atoms. As a result, a perfectly flat band emerges. However, despite having this completely flat band, the basic dice lattice has a zero Berry phase, meaning the system is topologically trivial \cite{PhysRevB.98.075422}.

Recently, efforts have been made to combine the effects of geometric moir\'e band folding with quantum interference. In the context of the $\alpha-T_3$ lattice, a relative twist between two layers generates a highly degenerate flat band. However, it has been theoretically demonstrated that aligning this twisted system with hexagonal boron nitride (hBN) substrates breaks the sublattice symmetry by introducing a staggered mass term. This crucial symmetry breaking lifts the large degeneracy of the dice limit and transforms the intrinsically trivial flat bands into emergent, non-degenerate topological subbands \cite{pyjd-jyrc}. 

Beyond the conventional Hermitian systems, non-Hermitian physics has recently emerged as a new way to engineer exotic band structures \cite{Okuma2023, PhysRevX.9.041015, PhysRevLett.115.200402, RevModPhys.93.015005}. By introducing asymmetric (or nonreciprocal) hopping of electrons, non-Hermiticity creates entirely new topological phases and unique spectral features such as exceptional points \cite{PhysRevLett.77.570, PhysRevLett.120.146402}. Furthermore, this non-reciprocity leads to the breakdown of the conventional bulk-boundary correspondence (BBC). As a consequence, a striking phenomenon known as the non-Hermitian skin effect (NHSE) emerges, where a macroscopic number of bulk eigenstates become exponentially localized at the open boundaries of the system \cite{PhysRevLett.121.086803, PhysRevLett.121.136802}. To accurately restore the BBC and define topological invariants in the presence of the NHSE, the standard Bloch band theory is fundamentally modified by replacing real wave vectors with complex momenta, giving rise to the generalized Brillouin zone (GBZ) \cite{PhysRevLett.123.066404, PhysRevB.101.195147}. Very recently, the broader concepts of non-Hermiticity have also been applied to moir\'e materials, specifically the tBG. Initial studies showed that introducing nonreciprocal hopping in tBG splits the conventional (Hermitian) magic angle into distinct exceptional magic angles (EMAs), which are associated with exceptional degeneracies and lead to a spectral gap closure \cite{dl59-vl7v}. Building on this, further investigations demonstrated that aligning this non-Hermitian tBG with an hBN substrate breaks the required rotational symmetries, opening a band gap and thereby creating a non-Hermitian valley-Hall insulating phase \cite{dfzm-hj41}. This revealed a rich interplay between EMAs, non-Hermiticity, and topological phase transitions.

Despite these recent advancements in non-Hermitian moir\'e systems, the impact of nonreciprocal hopping on the flat bands of the twisted bilayer $\alpha-T_3$ lattice remains unexplored. In this work, we investigate a non-Hermitian twisted bilayer $\alpha-T_3$ lattice by introducing Hatano-Nelson-type asymmetric hopping. We reveal that non-Hermiticity splits the single magic-angle condition of the Hermitian limit into three distinct non-Hermitian magic angles (NHMAs). Crucially, unlike the EMAs recently reported in tBG, these NHMAs host perfectly isolated flat bands where the complex bandwidth entirely vanishes without inducing any spectral singularities or band touchings. Furthermore, by analyzing the complex eigenspectra of these flat bands, we demonstrate that increasing the nonreciprocity causes the scattered energy eigenvalues to coalesce into distinct, closed loop-like structures in the complex energy plane, signifying a nontrivial point-gap topology. Additionally, we characterize the topological phase transitions of the system, revealing that strong nonreciprocal hopping drives the gap-closing boundaries to approach one another and ultimately merge. This pairwise topological charge annihilation manifests as a trivial gap closing, completely suppressing the robust higher-order topological phases.

The remainder of this paper is organized as follows: In Sec.~\ref{sec2}, we introduce the continuum Hamiltonian for the non-Hermitian twisted bilayer $\alpha-T_3$ lattice. In Sec.~\ref{sec3}, we discuss the emergence of three distinct NHMAs and analyze the flat-band conditions across the phase diagram. In Sec.~\ref{sec4}, we examine the complex eigenspectra of these isolated flat bands and the formation of spectral loops. In Sec.~\ref{sec5}, we explore the topological phase transitions driven by non-Hermiticity and the associated topological charge annihilation. Finally, our key findings are summarized in Sec.~\ref{sec6}.

\section{Model Hamiltonian and formalism}
\label{sec2}
\begin{figure}[t]
\centering
\begin{tabular}{c}
\hspace{-1cm}
\includegraphics[width=0.28\textwidth]{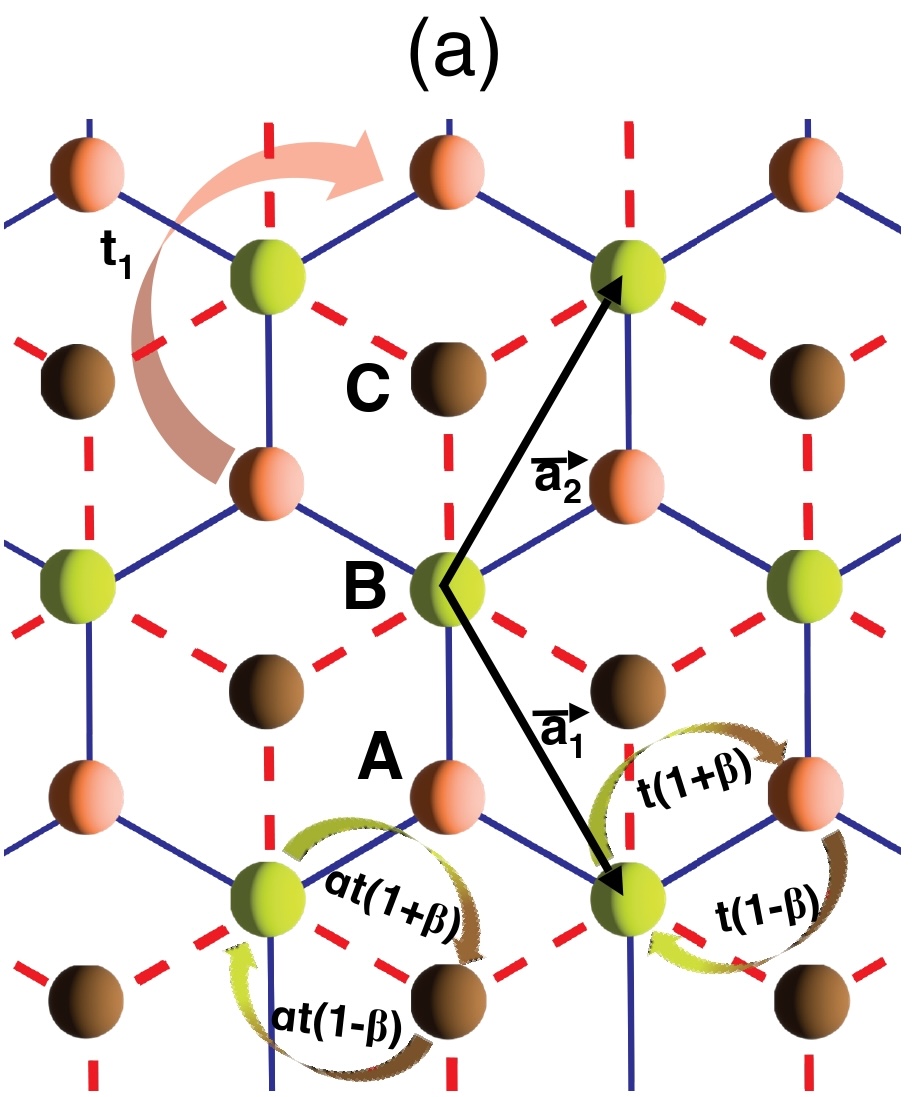} \\
\hspace{-0.5cm}
\includegraphics[width=0.5\textwidth]{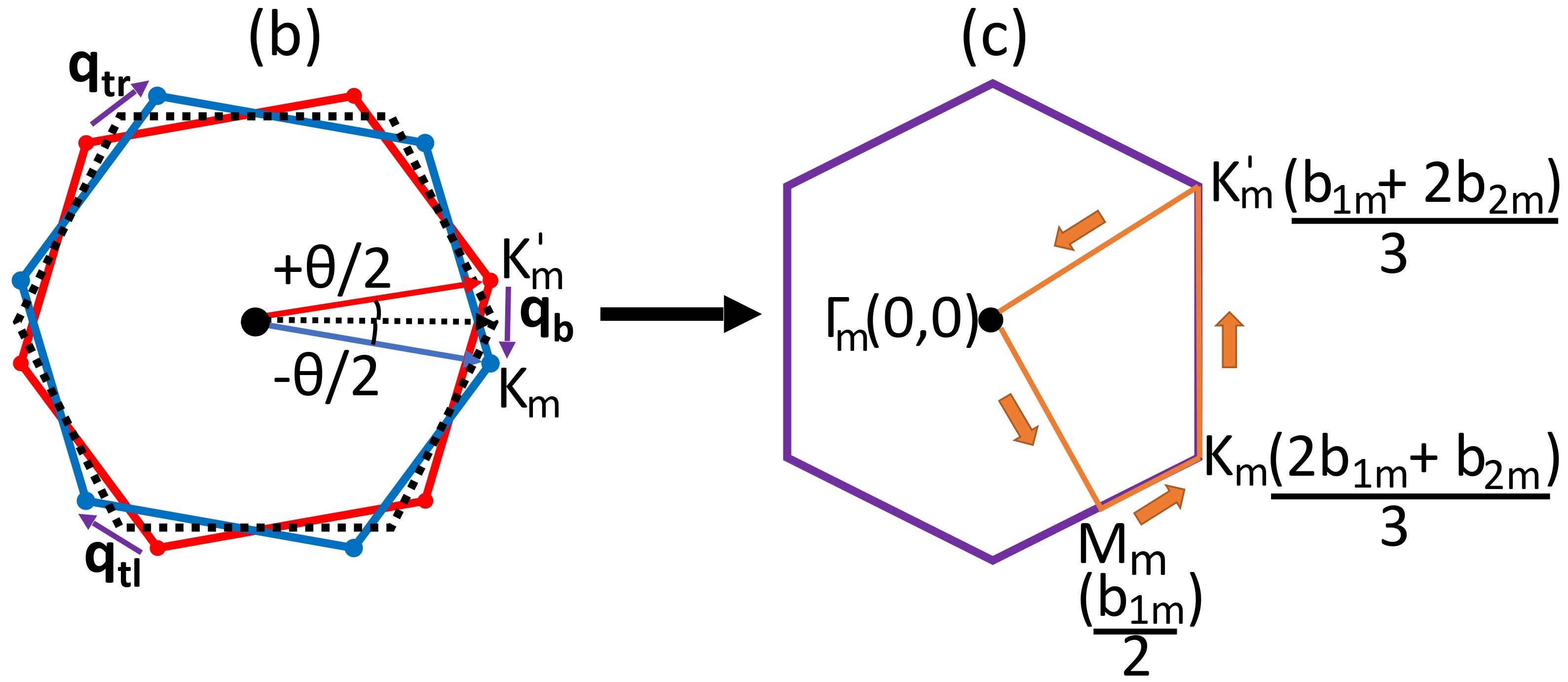} \\
\end{tabular}
\caption{(a) Schematic representation of the $\alpha$–$T_3$ lattice, showing the three sublattices A, B, and C. The non‑Hermitian effect is introduced through asymmetric hopping amplitudes, $t(1+\beta)$ and $t(1-\beta)$, between the A and B sites. Here, $t_1$ represents the next nearest‑neighbor hopping.(b) Brillouin zones of the top layer (blue) and bottom layer (red), rotated by $-\theta/2$ and $+\theta/2$, respectively, relative to the unrotated Brillouin zone (black dotted hexagon). The vectors $\mathbf{q}b$, $\mathbf{q}{tr}$, and $\mathbf{q}_{tl}$ indicate the momentum‑transfer vectors connecting the shifted Dirac points of the two layers.(c) The corresponding mBZ highlighting the high‑symmetry points ($\Gamma_m$, $M_m$, $K_m$, $K'_m$) and their coordinates. The arrows indicate the high‑symmetry momentum path used to calculate the energy bands.}
\label{fig:graphene}
\end{figure}

We consider a twisted bilayer $\alpha-T_3$ lattice in which the top (layer1) and bottom (layer2) sheets are rotated by $-\theta/2$ and $\theta/2$, respectively, as shown in Fig.~\ref{fig:graphene}(b). The intra-layer geometry, including the A, B, and C sublattices along with the nonreciprocal hopping parameters, is schematically illustrated in Fig.~\ref{fig:graphene}(a). Restricting to a single valley and neglecting intervalley scattering, the continuum Hamiltonian following the Bistritzer-MacDonald construction is \cite{Bistritzer2011}
\begin{equation}
H(\theta)=
\begin{pmatrix}
H_0^{(1)}\!\left(-\frac{\theta}{2}\right) & T \\
T^\dagger & H_0^{(2)}\!\left(+\frac{\theta}{2}\right)
\end{pmatrix}.
\end{equation}
Here $H_0^{(\ell)}$ denotes the intralayer Hamiltonian of layer $\ell=1,2$, and $T$ describes the interlayer tunneling processes.

The low-energy Hamiltonian of a single $\alpha-T_3$ layer expanded around the rotated Dirac point $K_\theta$ reads
\begin{equation}
\begin{aligned}
H_{K_\theta}(\mathbf{q})
&=
v_F
\begin{pmatrix}
0 & (1+\beta) q_- & 0 \\
(1-\beta) q_+ & 0 & \alpha (1+\beta) q_- \\
0 & \alpha (1-\beta) q_+ & 0
\end{pmatrix} \\
&\quad +\, t_2 q^2 I_3 + M S_z,
\end{aligned}
\end{equation}
with $q_\pm = q_x \pm i q_y$, $q^2 = q_x^2 + q_y^2$, $I_3 = \mathrm{diag}(1,1,1)$, and $S_z = \mathrm{diag}(1,0,-1)$.

The valley degree of freedom is incorporated through a valley index $\xi=\pm1$, which is implemented via $q_x \rightarrow \xi q_x$. Here, $\xi=+1$ corresponds to the $K$ valley and $\xi=-1$ to the $K'$ valley of the original $\alpha-T_3$ lattice, which are related to each other by time-reversal symmetry. To prevent notational ambiguity, we explicitly denote the high-symmetry corners of the emergent moir\'e Brillouin zone (mBZ) as $K_m$ and $K_m'$ [see Fig.~\ref{fig:graphene}(b,c)]. Restricting our focus to the $\xi=+1$ valley, the momentum $\mathbf{q} = \mathbf{k} - \mathbf{K}_\theta$ is measured relative to the rotated Dirac point $\mathbf{K}_\theta$.

The parameter $\alpha \in [0,1]$ interpolates between the graphene ($\alpha=0$) and dice ($\alpha=1$) limits. The quadratic term proportional to $t_2$ originates from next-nearest-neighbor hopping and regularizes the otherwise perfectly flat band of the dice lattice. The term $M S_z$ models the staggered mass induced by aligned hBN substrates. Non-Hermiticity is controlled by the parameter $\beta$, which introduces an asymmetry between the forward and backward Dirac hopping amplitudes in the off-diagonal matrix elements of $H_{K_\theta}(\mathbf{q})$, as shown in Fig. \ref{fig:graphene} (a). For $\beta \neq 0$, the hopping amplitudes along mutually opposite directions become unequal, leading to a nonreciprocal dispersion. This structure is analogous to Hatano-Nelson type non-Hermiticity, where asymmetric hopping generates complex spectra, while preserving the underlying lattice structure. In the Hermitian limit $\beta=0$, the conventional $\alpha-T_3$ Dirac Hamiltonian is recovered \cite{pyjd-jyrc}.

The interlayer tunneling Hamiltonian $T(\mathbf{q},\tilde{\mathbf{q}})$ connects the nearest Dirac points of the two layers in the mBZ via three principal momentum-transfer vectors. The explicit derivation of the moir\'e reciprocal lattice vectors, the momentum transfer vectors, and the corresponding $3 \times 3$ tunneling matrices $T_{\mathbf{q}_j}$ parameterized by the hopping amplitudes $w_1, w_2,$ and $w_3$ is presented in Appendix~\ref{app:tunneling}.

To diagonalize the full Hamiltonian, we expand the wavefunctions in a plane-wave basis constructed as the tensor product of layer, sublattice, and moir\'e reciprocal lattice degrees of freedom. The basis states are written as $\left| \ell, X; \mathbf{G}_{mn} \right\rangle,$ with $\ell$ labelling the layers, $X\in\{A,B,C\}$ denoting the sublattice index, and $\mathbf{G}_{mn}$ representing reciprocal lattice vectors of the moir\'e superlattice. For a plane-wave cutoff $N$, the total number of moir\'e reciprocal lattice points is $(2N+1)^2$. Including two layers and three sublattices per layer, the total Hilbert space dimension becomes $\mathcal{D} = 6(2N+1)^2$. In our calculations, we use $N=5$, resulting in a matrix dimension of $726$. The non-Hermitian Hamiltonian is then diagonalized numerically at each crystal momentum $\mathbf{k}$ in the mBZ.

\section{Results : Non-Hermitian Magic Angles}
\label{sec3}
\begin{figure*}[!htbp]
\centering

\begin{tabular}{cc}
\includegraphics[width=0.39\textwidth]{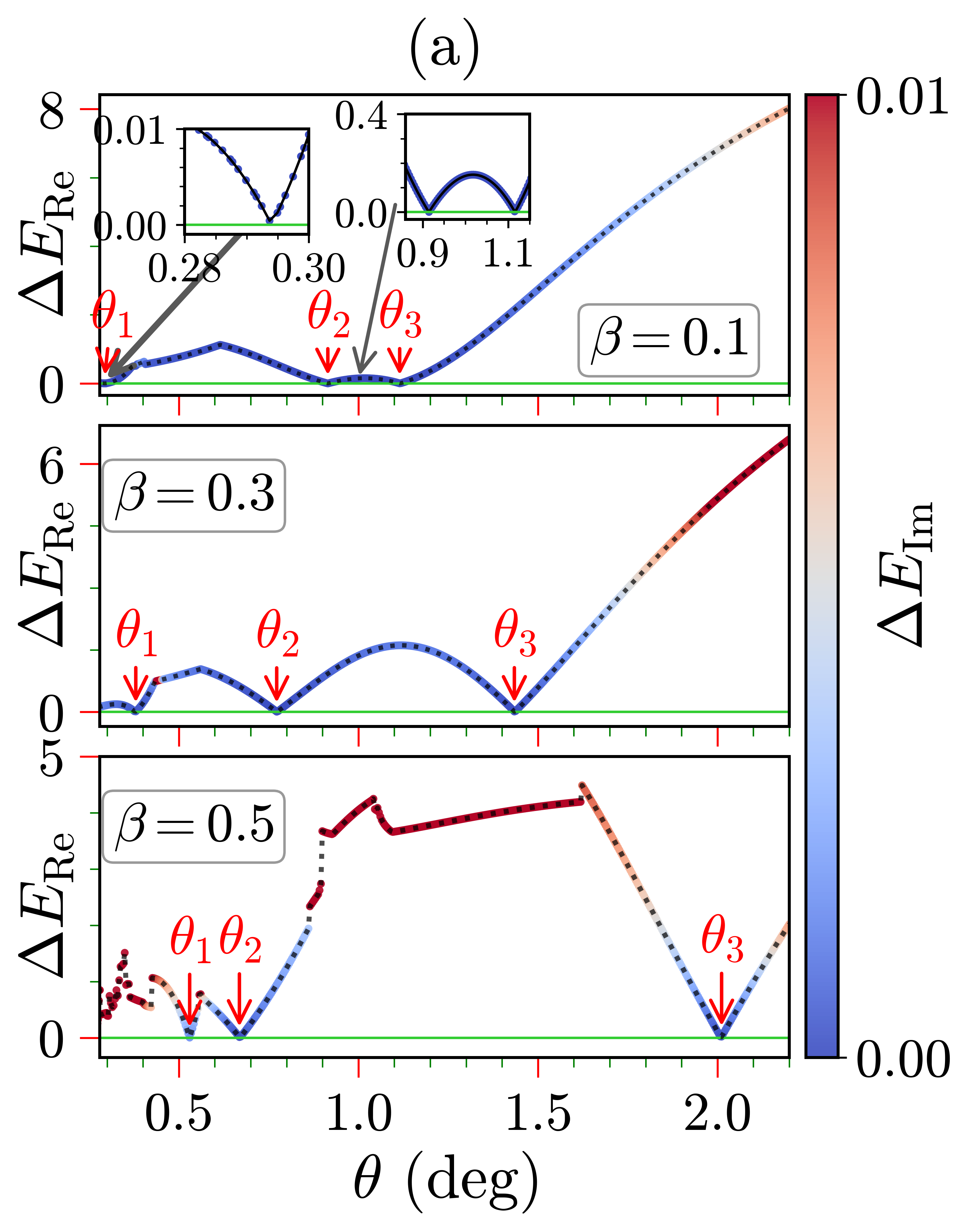} &
\includegraphics[width=0.4\textwidth]{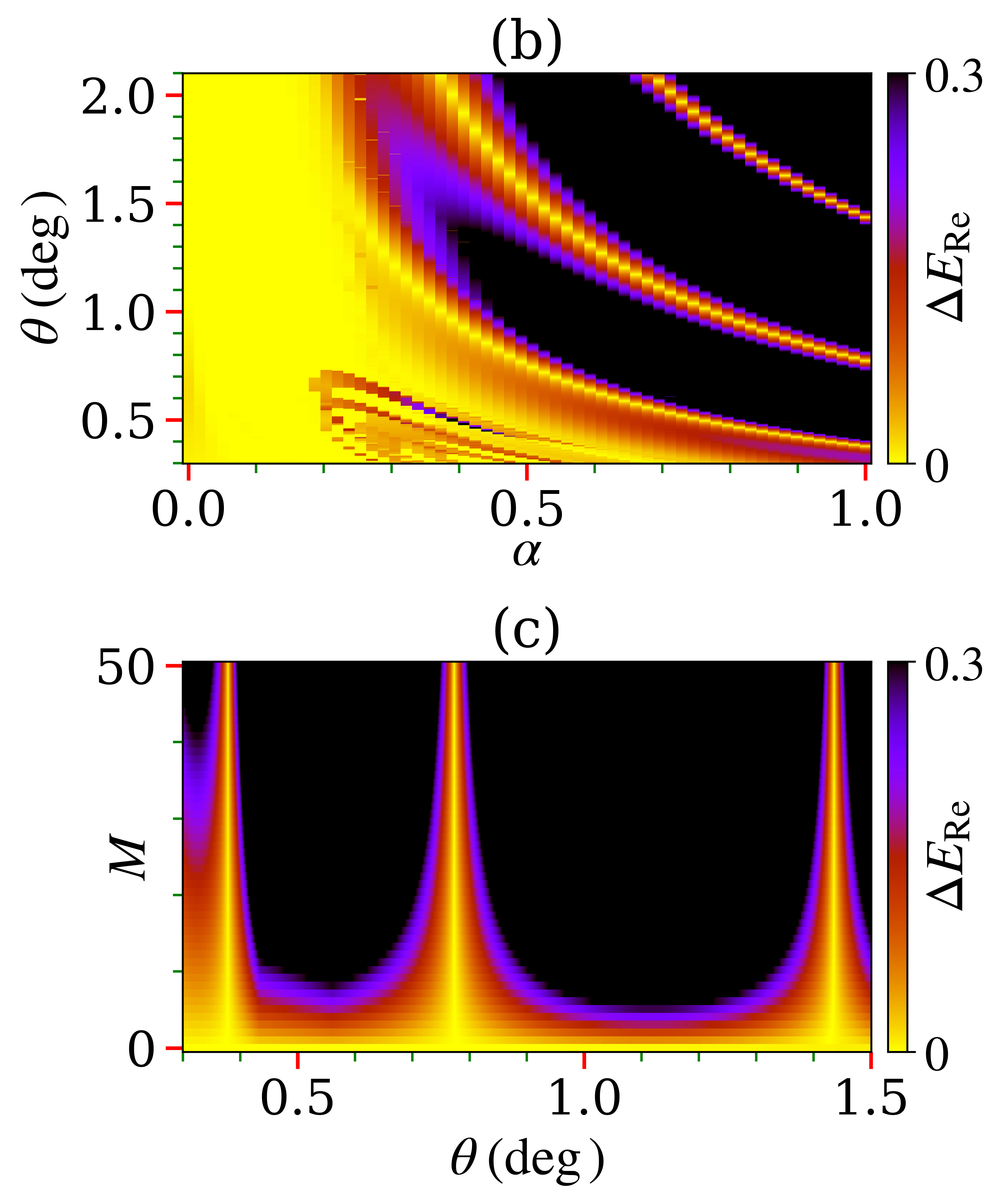}
\end{tabular}
\caption{(a) Real‑part bandwidth ($\Delta E_{\text{Re}}$) of the isolated band near $E_0 = -17$~meV as a function of the twist angle $\theta$ for selected non‑Hermitian strengths $\beta = 0.1$ (top panel), $0.3$ (middle panel), and $0.5$ (bottom panel) in the dice limit ($\alpha = 1$). The color scale encodes the corresponding imaginary‑part bandwidth ($\Delta E_{\mathrm{Im}}$). The NHMAs $\theta_1$, $\theta_2$, and $\theta_3$, at which $\Delta E_{\mathrm{Re}}$ and $\Delta E_{\mathrm{Im}}$ vanish within numerical precision, are marked by red arrows. These points demonstrate the splitting of the flat‑band condition into multiple NHMAs. The inset in the top panel shows a zoomed‑in view near the NHMAs for $\beta = 0.1$. (b) Phase diagram of the real‑part bandwidth $\Delta E_{\text{Re}}$ in the $(\alpha, \theta)$ plane for a fixed non‑Hermitian parameter $\beta = 0.3$. The bright yellow regions indicate perfectly flat bands ($\Delta E_{\text{Re}} \to 0$), whereas darker regions represent highly dispersive bands. The phase diagram highlights the transition from three distinct NHMAs near the dice limit ($\alpha \to 1$) to a broad, twist‑angle‑independent localized flat band in the graphene limit ($\alpha \to 0$). (c) Phase diagram of $\Delta E_{\text{Re}}$ in the $(M, \theta)$ plane for $\beta = 0.3$ and $\alpha = 1$. The vertical yellow strips indicate that the locations of the three NHMAs are robust and entirely independent of the staggered mass term $M$ induced by the aligned substrate.}
\label{fig:bandwidth_theta}

\end{figure*}

In the Hermitian limit ($\beta=0$), the twisted bilayer $\alpha-T_3$ lattice with a staggered mass $M$ is known to host nearly flat bands centered at $E=\pm M$ at the conventional magic angle $\theta\simeq 1.08^\circ$ in the dice limit ($\alpha=1$). To maintain consistency with these established Hermitian results, we set the staggered mass term to $M = 17~\text{meV}$ and specifically focus our analysis on the isolated band centered near $E_0 = -17~\text{meV}$ \cite{PhysRevResearch.1.033126, pyjd-jyrc}. To identify flat-band formation in the non-Hermitian regime, where the energy spectrum is generally complex, we compute both the real and imaginary parts of the bandwidth for this isolated band over the mBZ, defined as follows:
\begin{subequations}
\label{eq:bandwidths}
\begin{align}
\Delta E_{\text{Re}}(\theta) &= \max_{\mathbf{k}\in\mathrm{mBZ}} \mathrm{Re}\,E_n(\mathbf{k}) - \min_{\mathbf{k}\in\mathrm{mBZ}} \mathrm{Re}\,E_n(\mathbf{k}), \label{eq:bandwidth_re} \\
\Delta E_{\text{Im}}(\theta) &= \max_{\mathbf{k}\in\mathrm{mBZ}} \mathrm{Im}\,E_n(\mathbf{k}) - \min_{\mathbf{k}\in\mathrm{mBZ}} \mathrm{Im}\,E_n(\mathbf{k}), \label{eq:bandwidth_im}
\end{align}
\end{subequations}
where $E_n(\mathbf{k})$ denotes the complex eigenvalue selected at each momentum $\mathbf{k}$ such that its real part is closest to $E_0=-17$~meV. 

Figure~\ref{fig:bandwidth_theta}(a) illustrates the evolution of the real-part of the bandwidth ($\Delta E_{\text{Re}}(\theta)$) for selected values of the non-Hermitian parameter, namely $\beta = 0.1,\, 0.3$, and $0.5$. Crucially, the corresponding imaginary-part $\Delta E_{\text{Im}}(\theta)$ is simultaneously represented by the color map on the right side of these curves. In contrast to the single magic angle present in the Hermitian case, we observe that the inclusion of non-Hermiticity ($\beta\neq0$) splits the flat-band condition into three distinct minima. We denote the corresponding twist angles by $\theta_1, \theta_2,$ and $\theta_3$, and refer to them as non-Hermitian magic angles (NHMAs). At these specific points, both the real and imaginary bandwidths vanish within the numerical precision employed in our calcu;ations. These NHMAs are explicitly highlighted in Fig. \ref{fig:bandwidth_theta}(a) by red arrows. Furthermore, an inset provides a magnified view for $\beta = 0.1$ to clearly display the NHMAs.
With increasing $\beta$, the angular separation between $\theta_1$ and $\theta_2$ decreases, whereas the separation between $\theta_2$ and $\theta_3$ increases, indicating that asymmetric hopping modifies the flat-band condition. As shown by the color scale, the imaginary part of the bandwidth also becomes zero at the exact same twist angles where the real part is zero. This means that while increasing the non-Hermiticity shifts the locations of the magic angles, the bands formed at these new angles are completely flat.  Specifically, they are dispersionless in both their real and imaginary energy components.

To further explore the interplay between the lattice interpolation parameter $\alpha$ and the twist angle $\theta$, we present the phase diagram of the real-part bandwidth $\Delta E_{\text{Re}}$ in the $(\alpha, \theta)$ plane for a fixed non-Hermiticity $\beta = 0.3$, as shown in Fig.~\ref{fig:bandwidth_theta}(b). In this figure, the bright yellow regions indicate a completely flat band ($\Delta E_{\text{Re}} \to 0$), whereas the darker regions demonstrate the presence of highly dispersive bands. Near the dice limit ($\alpha \to 1$), the phase diagram clearly displays three distinct, sharply defined yellow spots, which correspond to the three separated NHMAs. As $\alpha$ assumes the intermediate values between $0$ and $1$, the bandwidth along these branches increases (indicated by the transition to darker colors), implying that the perfectly flat-band conditions are disrupted. This behavior is fully consistent with the Hermitian case, where the intermediate values of $\alpha$ are known to reduce the band flatness due to sub-optimal moir\'e destructive interference \cite{pyjd-jyrc}. As the system approaches the graphene limit ($\alpha \to 0$), the bandwidth approaches zero, resulting in a broad flat-band region (yellow) completely independent of the twist angle $\theta$. This phenomenon occurs because, at $\alpha = 0$, the C sublattice completely decouples from the A and B sublattices. Since the staggered mass term $M S_z$ assigns an on-site potential of $M = -17$~meV to the C sites, these decoupled states form a trivially flat and highly localized band exactly at $E_0 = -17$~meV. Therefore, the flat band at low values of $\alpha$ is a consequence of atomic localization rather than moir\'e physics. This clearly establishes that the emergence of exactly three distinct NHMAs near $\alpha=1$ is a unique interference effect of the non-Hermitian dice lattice, requiring the active coupling of all three sublattices. 

Having established the role of the lattice geometry, we now address the impact of the staggered mass, $M$. In standard moir\'e systems, such as tBG, the introduction of a staggered mass term via an aligned hBN substrate generally breaks the sublattice symmetry, which not only gaps the Dirac cones but also induce a dispersive character to the flat bands, thereby perturbing the perfect flat-band scenario \cite{PhysRevResearch.1.033126}. To investigate the specific effect of such substrate-induced mass on the non-Hermitian twisted bilayer $\alpha-T_3$ lattice, we present the phase diagram in the $(\theta, M)$ plane for a fixed non-Hermitian parameter $\beta = 0.3$ and $\alpha=1$ in Fig.~\ref{fig:bandwidth_theta}(c). Here, the color scale represents the real-part of the bandwidth ($\Delta E_{\text{Re}}$). Notably, the vertical bright yellow strips demonstrate that the locations of the three distinct NHMAs remain entirely independent of staggered mass $M$. This observationreveals that, unlike the conventional Hermitian moir\'e systems, the flat-band conditions here are governed predominantly by non-Hermiticity and quantum interference i.e. via the parameter $\alpha$. Consequently, the NHMAs in this system are highly robust against substrate-induced sublattice symmetry-breaking perturbations. In the following Sec.~\ref{sec4}, we present the complex eigenspectra to explicitly visualize the flat bands at these twist angles.

\section{Eigenspectra at NHMA}
\label{sec4}

To investigate the flat-band properties identified through the bandwidth analysis thoroughly, we divide our spectral analysis into two parts. First, we examine the band dispersion along the high-symmetry paths in momentum space. Second, we analyze the distribution of the complex eigenvalues over the entire mBZ to gain a comprehensive understanding of the overall spectral topology in the complex energy plane.

\subsection{Band Dispersion along High-Symmetry Paths}

\begin{figure}[htbp]
\centering
\begin{tabular}{cc}
\includegraphics[width=0.27\textwidth]{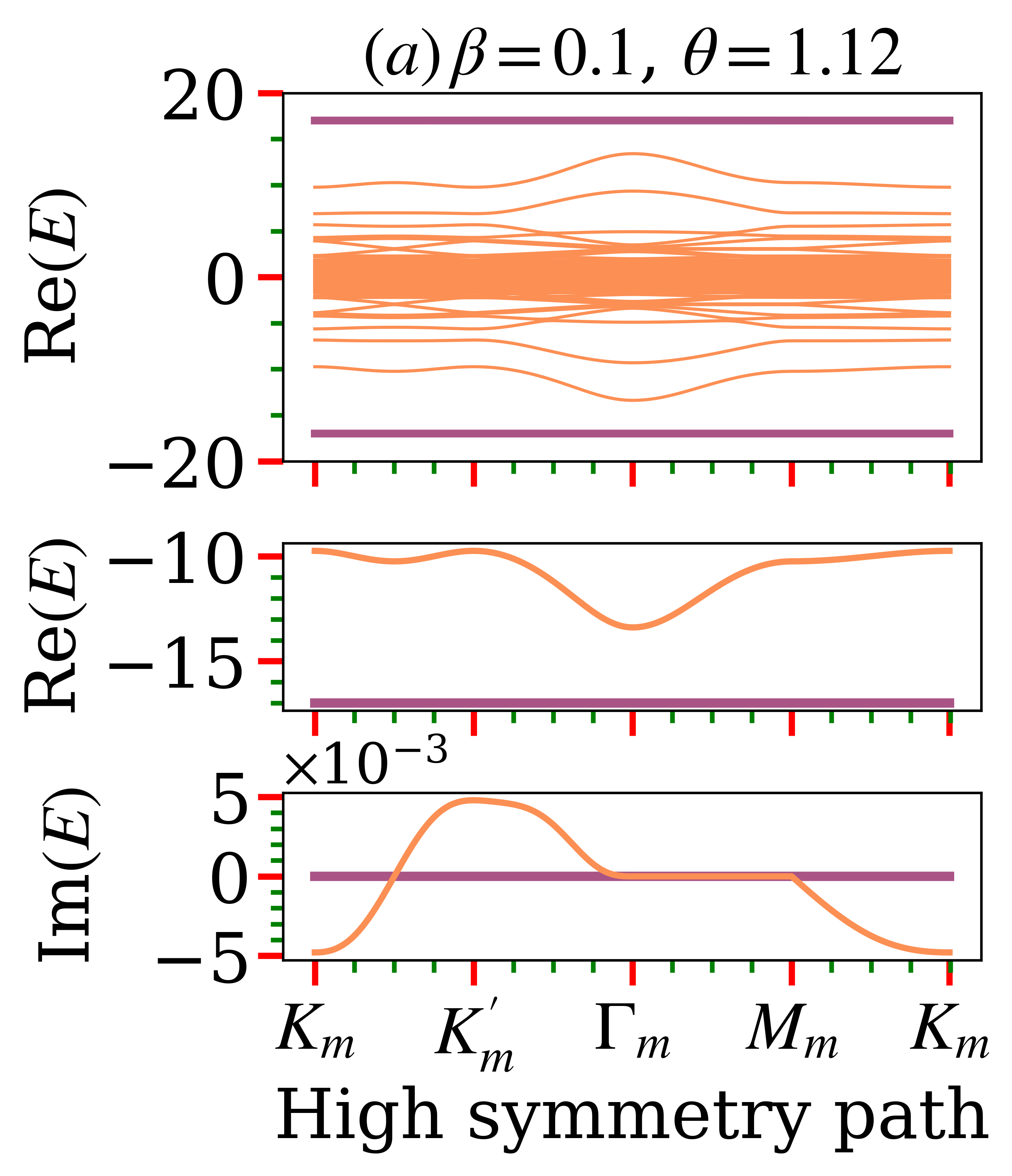} &
\includegraphics[width=0.195\textwidth]{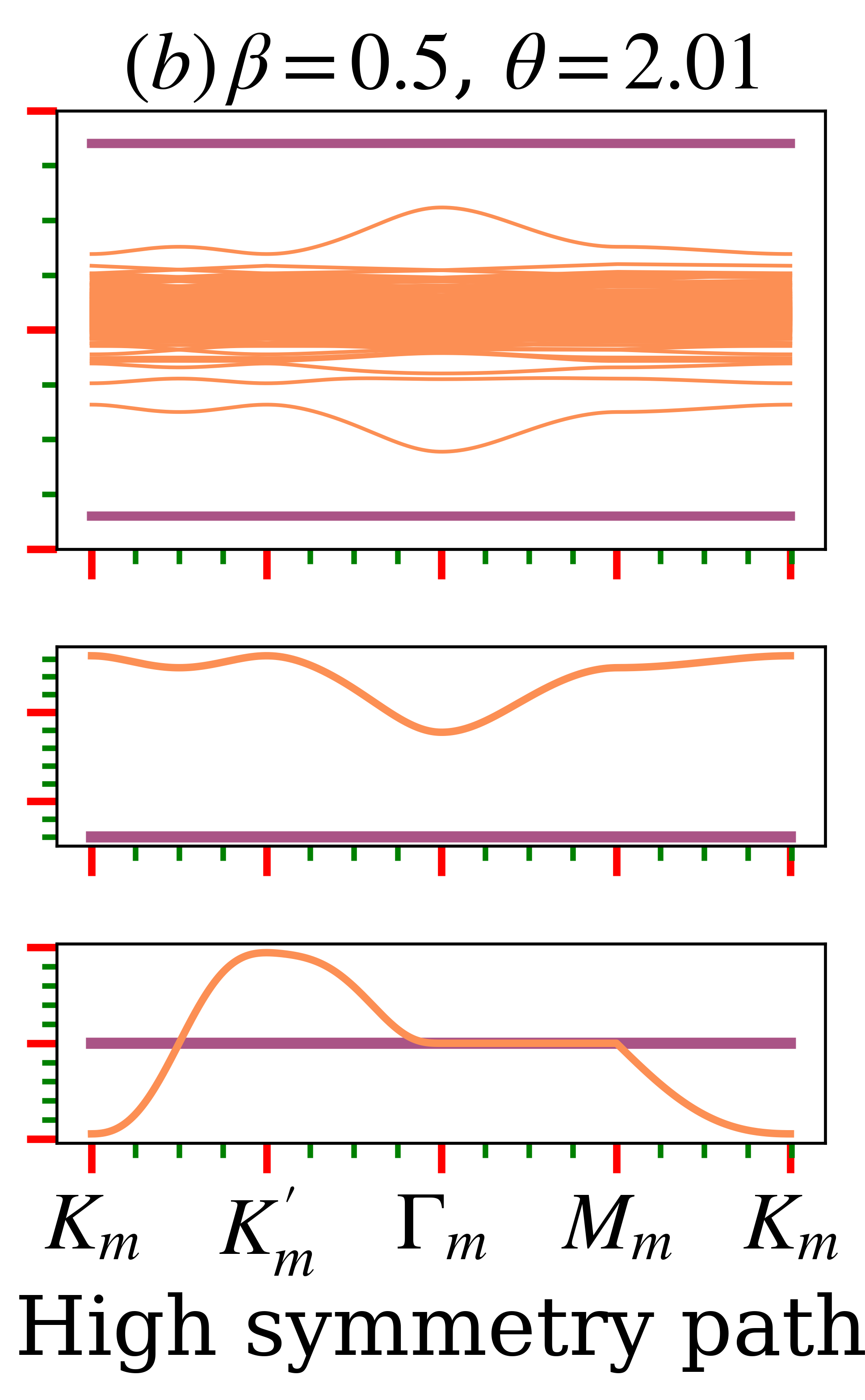} \\
\end{tabular}
\caption{
Complex energy bands evaluated along the high-symmetry path of the mBZ for two representative NHMAs at $(\beta=0.1, \theta=1.12^\circ)$ and $(\beta=0.5, \theta=2.01^\circ)$. The top panels show the full real-part spectrum,  featuring two isolated bands near $E \approx \pm 17$~meV. The middle and bottom panels display a zoomed-in views of the real and imaginary parts of the isolated band near $E_0=-17$~meV, respectively.
}
\label{fig:complex_spectra}
\end{figure}

To confirm the nature of the flat bands, we compute the full complex eigenspectrum of the continuum Hamiltonian along the high-symmetry path $K_m \rightarrow K'_m \rightarrow \Gamma_m \rightarrow M_m \rightarrow K_m$ of the mBZ, as shown in Fig.~\ref{fig:graphene}(c). Figure~\ref{fig:complex_spectra} illustrates the complex eigenspectra evaluated at two representative NHMAs, namely, $\theta=1.12^\circ$ for $\beta=0.1$, and $\theta=2.01^\circ$ for $\beta=0.5$. The real parts of the eigenvalues are presented in the upper panels, with a zoomed view of he isolated band near $E_0=-17$~meV shown in the middle panels, while the lower panels display the corresponding imaginary parts.

As shown in the top panels of Fig.~\ref{fig:complex_spectra}, two nearly dispersionless isolated bands are clearly observed in the real-part spectrum near $E \approx \pm 17$~meV (corresponding to the staggered mass terms $\pm M$). Consistent with our preceding bandwidth analysis, the middle panels provide a zoomed-in view of the lower flat band near $E_0 = -17$~meV. These zoomed panels confirm that this band remains perfectly flat and well-separated from the neighboring dispersive bands along the entire high-symmetry path, indicating complete spectral isolation at these NHMAs. Furthermore, the corresponding imaginary part of the eigenspectrum (bottom panels) demonstrates that the eigenvalue for this same band remains strictly momentum-independent and pinned precisely at zero along this entire high-symmetry path. Thus, the magic-angle condition here corresponds to a simultaneous and perfect suppression of the real and imaginary energies in the spectrum. 

We emphasize that, at the identified twist angles, the flat band remains spectrally isolated across the mBZ, and no band touching or eigenvector coalescence occurs. Therefore, unlike the EMAs recently reported in non-Hermitian tBG, the magic angles in the present twisted bilayer $\alpha-T_3$ system do not arise from the exceptional degeneracies. Instead, they correspond to perfectly isolated flat bands whose bandwidths (both the real and imaginary parts) vanish entirely without inducing any spectral singularities. This fundamental distinction highlights that the NHMAs in the non-Hermitian twisted bilayer $\alpha-T_3$ lattice represent a new class of flat-band conditions that are not driven by exceptional points, but rather by a unique interplay between moir\'e band folding and quantum interference that suppresses dispersion while maintaining spectral isolation. While the high-symmetry path analyses confirm the suppression of dispersion at the NHMAs, it does not provide a complete picture of the spectral distribution in the complex energy plane. To further elucidate the structure of the spectrum, we now examine the full complex eigenvalue distribution over the dense mBZ and analyze its dependence on the non-Hermitian parameter $\beta$ and twist angle $\theta$.

\subsection{Spectra in the Complex Energy Plane}

\begin{figure*}
\centering
\includegraphics[width=0.9\linewidth]{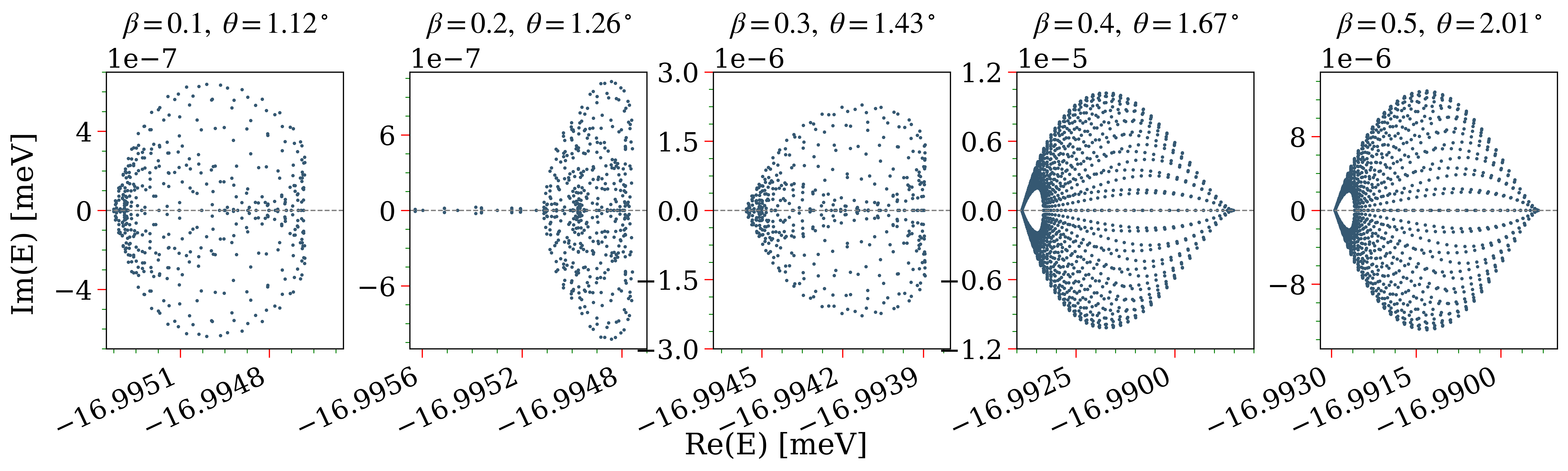}
\caption{Complex energy spectrum $(\mathrm{Re}\,E, \mathrm{Im}\,E)$ evaluated over a dense grid in the mBZ for the isolated band near $E_0 = -17$~meV. The panels show the evolution of the eigenvalues for increasing non-Hermitian parameter $\beta$ along the third magic-angle $\theta_3$. As the strength of the non-Hermitian parameter increases, the eigenvalues transform from scattered points to sharply defined, interconnected loop-like structures in the complex plane. Despite the visual prominence of these loops, their actual span remains microscopically small (up to $\sim10^{-5}$~meV), indicating that the flat band remains robustly pinned at $E_0 \approx -17$~meV. The formation of these closed spectral loops under periodic boundary conditions mathematically signifies a nontrivial point-gap topology, which is fundamentally tied to the NHSE.}

\label{fig:D2_biorth}
\end{figure*}

To further characterize the non-Hermitian flat band regime, we analyze the spectral structure by examining the real and imaginary parts of the eigenvalues across the full mBZ. For each parameter set $(\beta,\theta)$, the continuum Hamiltonian is diagonalized over a dense two-dimensional momentum grid, and the eigenvalues are mapped onto the complex energy plane defined by $\mathrm{Re}\,E$ and $\mathrm{Im}\,E$. In the present analysis, we focus exclusively on the isolated band centered in the vicinity of $E_0 \approx -17~\text{meV}$. Restricting the spectrum to a narrow energy window around this reference value allows us to strictly resolve the spectral features of the non-Hermitian flat band by avoiding the spectral overlap from the remote bands.

Figure~\ref{fig:D2_biorth} shows the resulting complex spectral distribution for a few values of $\beta$, evaluated at the corresponding third non-Hermitian magic-angle branch. The scattered plots display the complex eigenvalues computed over a dense mesh of momentum points $\mathbf{k}$ sampling the entire mBZ. As the non-Hermitian parameter $\beta$ increases, a striking structural evolution occurs in the complex energy plane. For smaller values of the non-Hermitian parameter, the eigenvalues form a diffuse, and somewhat scattered cloud-like distribution. However, as $\beta$ is increased, these scattered energies systematically condense into sharply defined, closed loop-like contours. This structural evolution reveals that larger asymmetric hopping amplitude induces a minute spectral splitting of the adjacent dispersive states, while the target flat band remains robustly pinned exactly at the singular point near $E_0 \approx -17$~meV. Notably, while these spectral loops are visually pronounced, their actual span in the complex energy plane is microscopically small (on the order of $10^{-6}$~meV). Despite this diminutive energy scale, the formation of closed eigenvalue loops under periodic boundary conditions mathematically signifies a non-trivial point-gap topology \cite{PhysRevX.9.041015, RevModPhys.93.015005}. In non-Hermitian systems, this bulk spectral feature is fundamentally tied to the NHSE, strongly suggesting that the corresponding eigenstates would macroscopically localize at the system boundaries under OBC \cite{PhysRevLett.121.086803, PhysRevLett.124.086801}. 

\begin{figure}[htbp]
\centering
\includegraphics[width=1\linewidth]{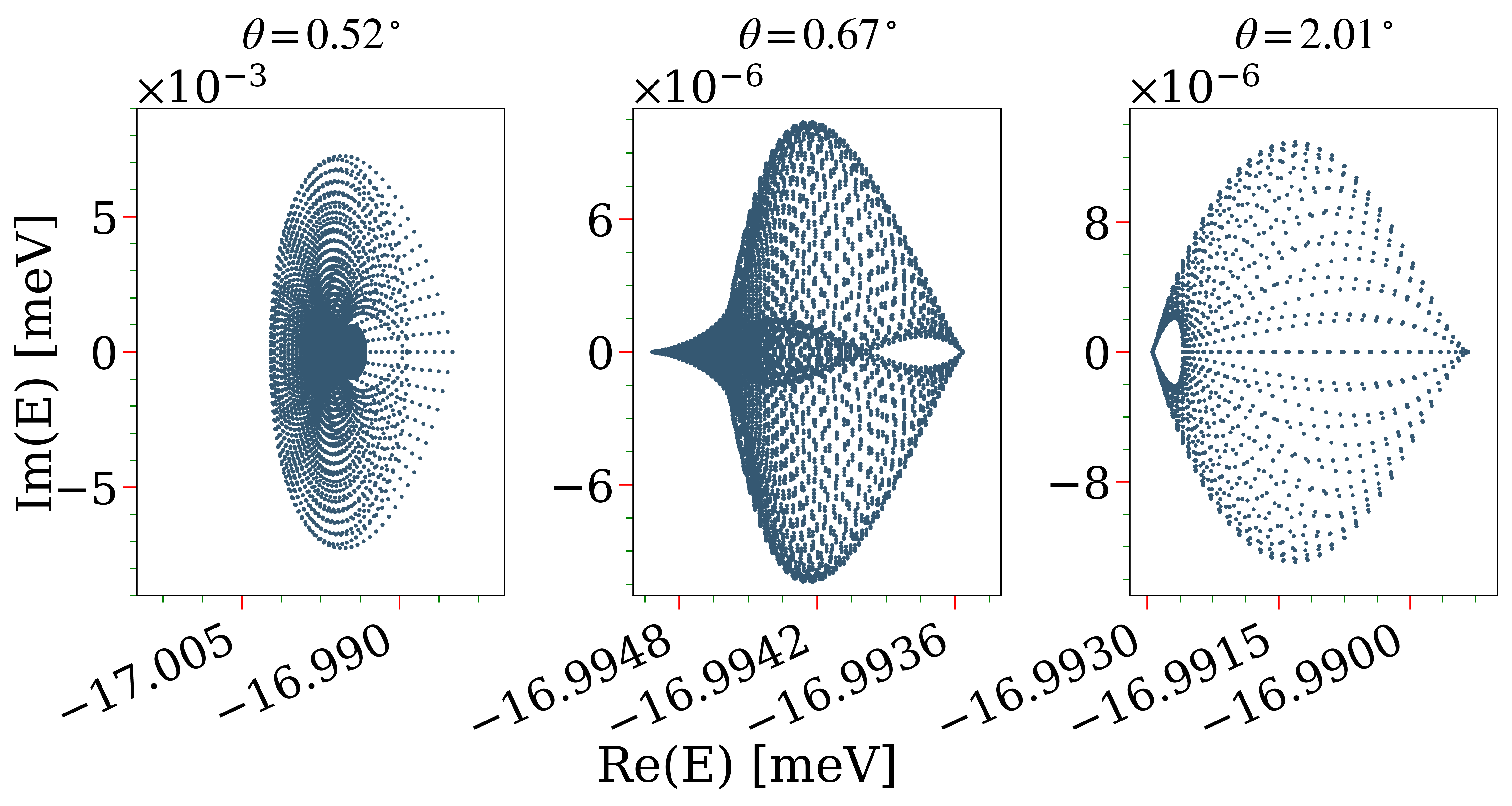}
\caption{Complex spectral distribution near $E_0 \simeq -17$~meV for a fixed non-Hermiticity strength $\beta=0.5$ evaluated at the three distinct NHMAs: $\theta_1=0.52^\circ$, $\theta_2=0.67^\circ$, and $\theta_3=2.01^\circ$. Although all three angles host perfectly flat bands, the geometric shape and energy span of the corresponding spectral loops vary significantly.}
\label{fig:D2_biorth_beta_0_5}
\end{figure}

Furthermore, to disentangle the respective roles of non-Hermiticity and the twist angle, we fix the non-Hermiticity strength at $\beta=0.5$ and examine how the complex spectral structure evolves along the three distinct non-Hermitian magic-angle branches. Figure~\ref{fig:D2_biorth_beta_0_5} displays the complex energy spectra evaluated at the three corresponding NHMAs, namely, $\theta_1=0.52^\circ$, $\theta_2=0.67^\circ$, and $\theta_3=2.01^\circ$. For all three twist angles, the eigenvalues fundamentally organize into dense, nested loop-like structures in the complex plane, confirming that non-reciprocal Dirac hopping generically reconstructs the flat-band spectrum. However, the exact geometric shape, symmetry, and spanning area of these complex bands differ markedly among the three cases. At the smallest twist angle ($\theta_1=0.52^\circ$), the spectrum spans an asymmetric, densely packed region. As the twist angle increases to $0.67^\circ$, and subsequently to $2.01^\circ$, the spectral loops undergo significant morphing, adopting sharply defined, symmetric spindle-like profiles. Notably, the microscopic energy scale of these complex deformations varies by several orders of magnitude across different twist angles, from an order of $10^{-3}$~meV at $0.52^\circ$ down to $10^{-6}$~meV at $0.67^\circ$, despite the non-Hermitian parameter $\beta$ being constant. Throughout these structural and scaling changes, the isolated flat band near $E_0 \simeq -17$~meV remains perfectly preserved. These observations demonstrate that the complex spectral structure in the non-Hermitian twisted $\alpha-T_3$ system is governed by a critical interplay between the non-Hermitian strength $\beta$ and the twist angle $\theta$. Furthermore, to establish the macroscopic presence of the NHSE within the physically relevant moir\'e continuum, a comparative spectral analysis over an extended low-energy window is provided in Appendix~\ref{appendixb}.

\section{Topological Phase Transitions}
\label{sec5}
\begin{figure}[htbp]
\centering
\includegraphics[width=1\columnwidth]{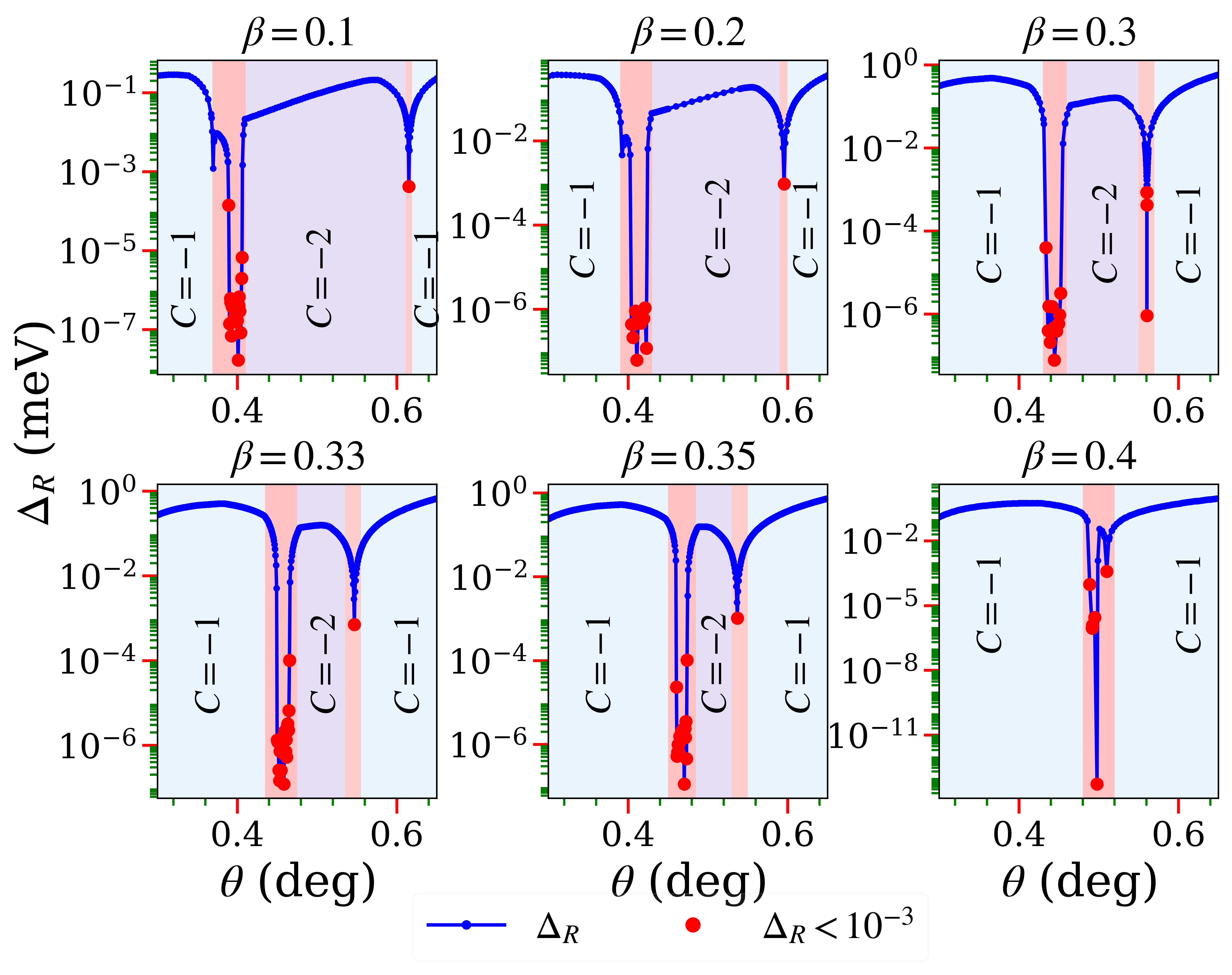}
\caption{Evolution of the real part of the direct band gap ($\Delta_R$) and the biorthogonal Chern number ($C$) as a function of the twist angle $\theta$ for varying non-Hermitian parameters $\beta$. The solid blue curves represent $\Delta_R$, while the red dots highlight the critical gap-closing points where $\Delta_R < 10^{-3}$~meV. The light-blue shaded regions denote the topological phases characterized by $C = -1$, whereas the purple shaded region corresponds to the $C = -2$ phase. The light-red vertical shaded bands indicate the gap-closing regions. As $\beta$ increases from $0.1$ to $0.4$, the two gap-closing boundaries migrate toward each other, progressively shrinking the twist-angle window of the $C = -2$ phase. At $\beta = 0.4$, these boundaries completely merge.}
\label{fig:Chern}
\end{figure}

To properly characterize the topological phases in a non-Hermitian twisted bilayer $\alpha-T_3$ lattice, we investigate the topological invariant that is accompanied by spectral gap closing transitions. The existence of a bulk gap-closing transition is a primary indicator of a topological phase transition. Because our primary focus is the topological evolution of the target flat band pinned near $E_0 \simeq -17$~meV, we strictly evaluate the energy gap between this specific flat band and the nearest adjacent band. Thus, at each $\mathbf{k}$ in the mBZ, the complex eigenvalues are sorted by their absolute distance relative to $E_0$ along the real axis. The eigenenergy closest to $E_0$, denoted as $E_{j_1}(\mathbf{k})$, naturally corresponds to the flat band state, while the second closest, $E_{j_2}(\mathbf{k})$, represents the nearest adjacent dispersive band. Consequently, we define the global real part of the direct band gap $\Delta_R$ between these two states as \cite{dfzm-hj41}
\begin{equation}
    \Delta_R = \min_{\mathbf{k} \in \text{mBZ}} \left| \mathrm{Re}[E_{j_2}(\mathbf{k})] - \mathrm{Re}[E_{j_1}(\mathbf{k})] \right|.
\end{equation}
A zero (or a vanishingly small) $\Delta_R$ signifies a bulk band-touching event in the real energy spectrum. To establish the topological nature of the phases separated by these gapless regions, we calculate the non-Hermitian generalization of the Chern number \cite{PhysRevLett.120.146402, PhysRevLett.121.136802}. Due to the broken Hermiticity induced by the non-reciprocal hopping, the left ($\langle \psi_{n}^{L}(\mathbf{k}) |$) and right ($| \psi_{n}^{R}(\mathbf{k}) \rangle$) eigenvectors form a biorthogonal basis satisfying $\langle \psi_{n}^{L}(\mathbf{k}) | \psi_{n}^{R}(\mathbf{k}) \rangle = 1$. The non-Hermitian Chern number for a separable energy band $n$ is defined as \cite{Fukui2005}:
\begin{equation}
    C = \frac{1}{2\pi} \int_{\text{mBZ}} d^2\mathbf{k} \, \epsilon_{ij} \, i \langle \partial_{k_i} \psi_{n}^{L}(\mathbf{k}) | \partial_{k_j} \psi_{n}^{R}(\mathbf{k}) \rangle,
\end{equation}
where $\epsilon_{ij}$ is the 2D antisymmetric tensor.

In Fig.~\ref{fig:Chern}, we present the real part of the direct band gap $\Delta_R$ and the corresponding Chern numbers as a function of the twist angle $\theta$ for various strengths of the non-Hermitian parameter ($\beta \in [0.1, 0.4]$). As $\theta$ is varied, the system undergoes distinct band-touching events, marked by sharp dips where $\Delta_R$ effectively vanishes within narrow intervals of critical twist angles (highlighted by the light-red shaded regions). For the regime up to $\beta \le 0.35$, the biorthogonal Chern number undergoes abrupt and quantized jumps exactly across these critical gapless regions. Varying $\theta$ drives topological phase transitions from a phase characterized by $C = -1$ to a phase with a higher Chern number of $C = -2$, and subsequently back to $C = -1$. This precise correspondence between the bulk gap closures and the discrete changes in the topological invariant confirms the topological phase transitions occurring herein \cite{PhysRevLett.120.146402, PhysRevLett.121.136802}.

However, a closer examination of the gap-closing boundaries across the successive panels reveals a strong interplay between the non-Hermiticity strength $\beta$ and the topological phases. The width of the twist-angle window supporting the $C = -2$ phase is highly sensitive to the nonreciprocity in the hopping term. As $\beta$ increases from $0.1$ to $0.35$, the two critical gap-closing regions migrate toward each other, causing the $C = -2$ topological phase to progressively shrink. This suppression continues until $\beta = 0.4$, where the two critical boundaries completely merge. This phenomenon can be understood through the topological charge conservation and the pairwise annihilation of non-Hermitian degeneracies (exceptional points) \cite{PhysRevX.9.041015}, a mechanism physically analogous to the pairwise annihilation of Weyl nodes carrying opposite chiralities in a 3D semimetals \cite{PhysRevB.80.153412, RevModPhys.90.015001}. In non-Hermitian topological band theory, such exceptional gap-closing boundaries act as sources of Berry curvature in the non-Hermitian parameter space, carrying a topological charge equal to the discrete change in the Chern number ($\Delta C$) \cite{RevModPhys.82.3045, RevModPhys.93.015005}. For $\beta \le 0.35$, the two boundaries possess equal and opposite topological charges: $\Delta C = -1$ (for the transition from $C=-1$ to $-2$) and $\Delta C = +1$ (for the transition from $C=-2$ to $-1$). These boundaries converge at $\beta = 0.4$, which is when their net topological charge becomes zero ($Q_{\text{net}} = -1 + 1 = 0$). This results in the mutual annihilation of their topological charges, manifesting as a trivial gap closing. Consequently, the intermediate $C = -2$ phase is completely suppressed, and the system globally retains its initial $C = -1$ phase. This behavior clearly demonstrates that strong nonreciprocity is fundamentally detrimental to the stability of intermediate topological phases in the present system. This observation is consistent with the broader phenomenon of non‑Hermiticity‑induced topological suppression recently reported in twisted bilayer graphene \cite{dfzm-hj41}, confirming that enhanced nonreciprocal hopping undermines the topological robustness of such moir\'e systems. Crucially, the suppression of the $C = -2$ phase is not driven by non‑Hermiticity alone, but rather emerges from a interplay between the lattice geometry ($\alpha$) and the twist angle ($\theta$). Accordingly, to provide a complete picture of how the underlying lattice geometry influences the topological phase structure, we present the phase diagram of the biorthogonal Chern number in the $(\alpha, \theta)$ plane in Appendix~\ref{appendixc}.

\section{Summary}
\label{sec6}
In summary, we have investigated the band structure and topological phase transitions of a non-Hermitian twisted bilayer $\alpha-T_3$ lattice. In Sec. \ref{sec2}, we present the model Hamiltonian of the system, incorporating Hatano-Nelson-type nonreciprocal hopping and a staggered mass arising from an aligned hBN substrate to explore how non-Hermiticity modifies the flat-band physics. Building upon this framework, our bandwidth analyses in Sec. \ref{sec3} reveal that the single magic angle found in the Hermitian limit splits into three distinct NHMAs in the dice limit ($\alpha \to 1$). By analyzing the phase diagram in the $(\alpha, \theta)$ plane, we further demonstrated that these three distinct NHMAs arise from a unique interplay between moir\'e band folding and quantum interference specific to the non-Hermitian dice lattice ($\alpha \to 1$). As $\alpha$ decreases, these distinct flat-band branches gradually broaden and merge. In contrast, transitioning completely towards the graphene limit ($\alpha \to 0$) yields a broad, twist-angle-independent localized flat band driven purely by decoupling of the atom at the center of the hexagon (C-sublattice). Moreover, by examining the phase diagram in the $(\theta, M)$ plane, we establish that the precise locations of these NHMAs are robust and completely independent of the staggered mass induced by the alignment of the substrate.

Subsequently, in Sec.~\ref{sec4}, we examined the full complex energy spectrum of these isolated flat bands. We observed that increasing the nonreciprocal hopping parameter ($\beta$) forces the scattered complex eigenvalues to coalesce into sharp, closed loop-like structures. The formation of these spectral loops signifies a non-trivial point-gap topology, which inherently points toward the emergence of the NHSE effect under open boundary conditions. Finally, in Sec.~\ref{sec5}, we mapped the topological phase diagram of the system by tracking the real part of the direct band gap and computing the biorthogonal Chern numbers. For weak non-Hermiticity ($\beta \le 0.35$), the system exhibits standard topological band behavior, undergoing multiple phase transitions that clearly discern transition from an initial phase $C = -1$ to a $C = -2$ phase, and back to $C = -1$. However, as the non-Hermiticity is further increased to a value $\beta = 0.4$, the twist-angle window supporting the $C = -2$ phase progressively shrinks. The two gap-closing boundaries, which carry equal and opposite topological charges, are driven together until they merge and mutually annihilate. This pairwise annihilation results in a net topological charge of zero, manifesting as a trivial gap closing that completely suppresses the intermediate $C = -2$ phase. Our findings conclusively demonstrate that while weak non-Hermiticity can support topological transitions, strong nonreciprocal hopping fundamentally destroys the robust higher-order topological phases, forcing the system to globally retain its initial $C = -1$ phase.

\begin{acknowledgments}
S.G. acknowledges the Anusandhan National Research Foundation (ANRF), Govt. of India, for providing financial support through the National Post Doctoral Fellowship (NPDF) (File No. PDF/2025/004365). S.G. further acknowledge National Supercomputing Mission (NSM) for providing computing resources of ‘PARAM Kamrupa’ at IIT Guwahati, which is implemented by C-DAC and supported by the Ministry of Electronics and Information Technology (MeitY) and Department of Science and Technology (DST), Government of India. G.P. acknowledges financial support from the Ministry of Education (MoE), Government of India, through the research fellowship. S.L. acknowledges financial support from the MoE, Govt. of India, through the Prime Minister’s Research Fellowship (PMRF) scheme in May 2022. 
\end{acknowledgments}

\section*{Data Availability}
The data that support the findings of this article are not publicly available. The data are available from the authors upon reasonable request.

\appendix

\section{Details of the Interlayer Tunneling Hamiltonian}
\label{app:tunneling}

The interlayer Hamiltonian connects the nearest Dirac points of the two layers in the moir\'e Brillouin zone. The moir\'e reciprocal lattice vectors are given by
\begin{subequations}
\begin{align}
\mathbf{b}_1^m &=
\frac{8\pi \sin(\theta/2)}{3d}
\left(
\frac{1}{2}, -\frac{\sqrt{3}}{2}
\right),
\\
\mathbf{b}_2^m &=
\frac{8\pi \sin(\theta/2)}{3d}
\left(
\frac{1}{2}, \frac{\sqrt{3}}{2}
\right).
\end{align}
\end{subequations}

Let $\mathbf{K}_1$ and $\mathbf{K}_2$ denote the positions of the rotated Dirac points for layer1 and layer2, respectively, which corresponds to the $K_m$ and $K_m'$ corners of the mBZ. They are given by
\begin{subequations}
\label{eq:Dirac_points}
\begin{align}
\mathbf{K}_1 &=
\frac{8\pi \sin(\theta/2)}{3\sqrt{3}d}
\left(
-\frac{\sqrt{3}}{2}, -\frac{1}{2}
\right), \label{eq:K1}
\\
\mathbf{K}_2 &=
\frac{8\pi \sin(\theta/2)}{3\sqrt{3}d}
\left(
-\frac{\sqrt{3}}{2}, \frac{1}{2}
\right). \label{eq:K2}
\end{align}
\end{subequations}

The three principal momentum transfer vectors connecting the nearest Dirac points in the mBZ are defined as
\begin{subequations}
\label{eq:momentum_transfer}
\begin{align}
\mathbf{q}_b &=
\frac{8\pi \sin(\theta/2)}{3\sqrt{3}d}
(0,-1), \label{eq:qb}
\\
\mathbf{q}_{tr} &=
\frac{8\pi \sin(\theta/2)}{3\sqrt{3}d}
\left(
\frac{\sqrt{3}}{2}, \frac{1}{2}
\right), \label{eq:qtr}
\\
\mathbf{q}_{tl} &=
\frac{8\pi \sin(\theta/2)}{3\sqrt{3}d}
\left(
-\frac{\sqrt{3}}{2}, \frac{1}{2}
\right). \label{eq:qtl}
\end{align}
\end{subequations}

Retaining only these three dominant scattering processes, the momentum-space interlayer tunneling Hamiltonian takes the form
\begin{align}
T(\mathbf{q},\tilde{\mathbf{q}})
&=
T_{\mathbf{q}_b}\,
\delta_{\mathbf{q}-\tilde{\mathbf{q}}-\mathbf{q}_b}
+
T_{\mathbf{q}_{tr}}\,
\delta_{\mathbf{q}-\tilde{\mathbf{q}}-\mathbf{q}_{tr}}
+
T_{\mathbf{q}_{tl}}\,
\delta_{\mathbf{q}-\tilde{\mathbf{q}}-\mathbf{q}_{tl}} .
\end{align}

The corresponding tunneling matrices $T_{\mathbf{q}_j}$ in the $(A,B,C)$ sublattice basis are explicitly given by:
\begin{subequations}
\label{eq:tunneling_matrices}
\begin{align}
T_{\mathbf{q}_b} &=
\begin{pmatrix}
w_1 & w_2 & w_3 \\
w_2 & w_1 & w_2 \\
w_3 & w_2 & w_1
\end{pmatrix}, \label{eq:Tqb}
\\[6pt]
T_{\mathbf{q}_{tr}} &=
\begin{pmatrix}
w_1 e^{i\xi 2\pi/3} & w_2 & w_3 e^{-i\xi 2\pi/3} \\
w_2 e^{-i\xi 2\pi/3} & w_1 e^{i\xi 2\pi/3} & w_2 \\
w_3 & w_2 e^{-i\xi 2\pi/3} & w_1 e^{i\xi 2\pi/3}
\end{pmatrix}, \label{eq:Tqtr}
\\[6pt]
T_{\mathbf{q}_{tl}} &=
\begin{pmatrix}
w_1 e^{-i\xi 2\pi/3} & w_2 & w_3 e^{i\xi 2\pi/3} \\
w_2 e^{i\xi 2\pi/3} & w_1 e^{-i\xi 2\pi/3} & w_2 \\
w_3 & w_2 e^{i\xi 2\pi/3} & w_1 e^{-i\xi 2\pi/3}
\end{pmatrix}. \label{eq:Tqtl}
\end{align}
\end{subequations}
\noindent where $w_1$, $w_2$, and $w_3$ are the interlayer hopping amplitudes between the respective sublattices of the two layers.

\section{Complex Eigenspectra in the Extended Low-Energy Moir\'e Continuum}
\label{appendixb}
\begin{figure*}[t]
    \centering
    \includegraphics[width=0.95\linewidth]{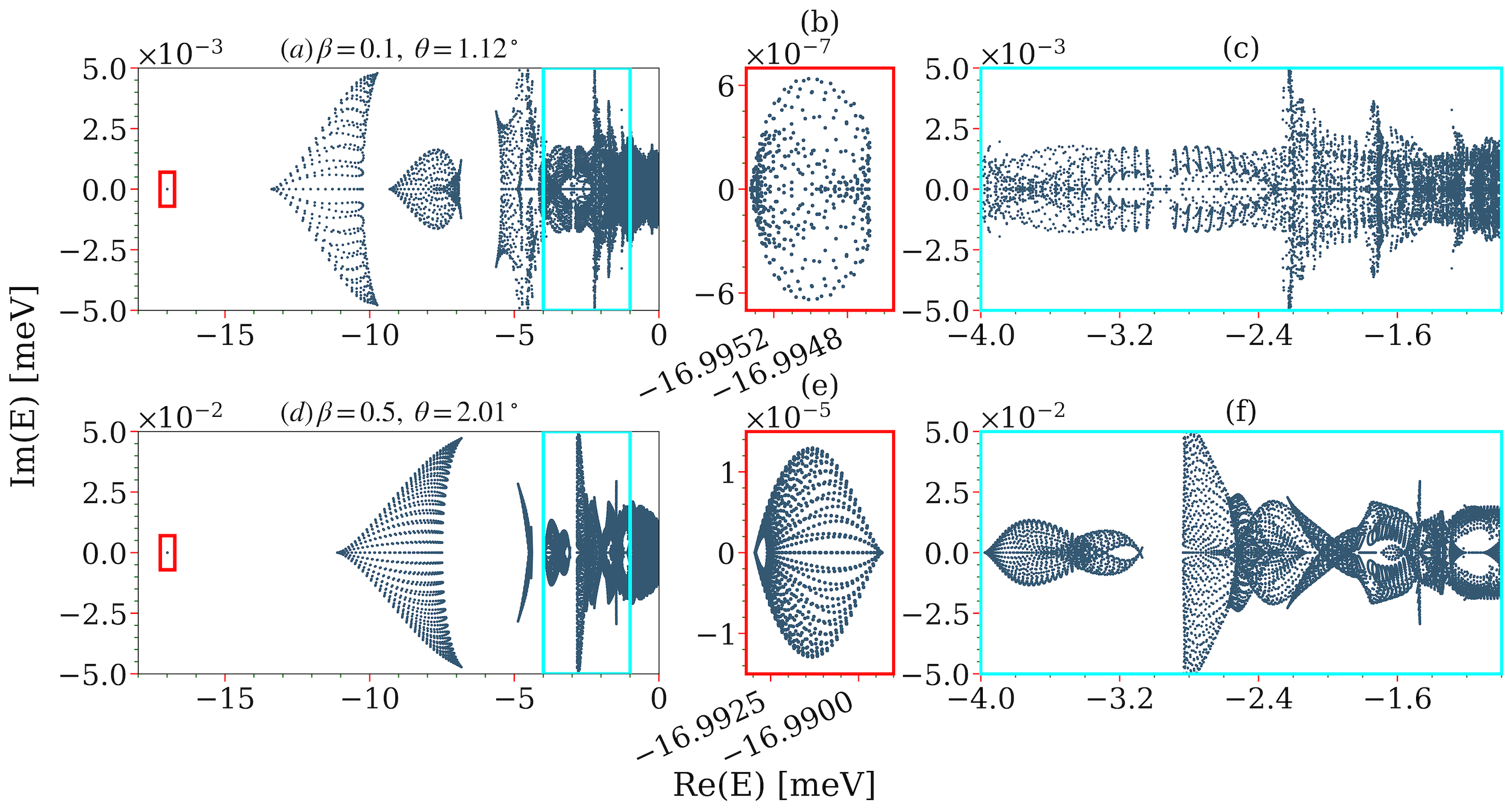}
   \caption{(a, d) Complex energy spectrum over an extended low-energy window for non-Hermitian parameter strengths $\beta = 0.1$ and $\beta = 0.5$, respectively. The red rectangles highlight the target isolated flat band near $E_0 \approx -17$~meV, while the cyan rectangles indicate the highly dispersive moir\'e bands. (b, e) Zoomed-in views of the isolated flat band. At $\beta = 0.1$, the eigenvalues appear as a scattered points, whereas at $\beta = 0.5$, they organize into microscopic closed loops. The highly constrained imaginary energy part ($\sim 10^{-5}$~meV) confirms the robust spectral pinning of the flat band, while the loop structure signifies its nontrivial point-gap topology. (c, f) Zoomed-in views of the dispersive moir\'e bands. Here, the spectrum undergoes a macroscopic transformation from a scattered points of eigenvalues at $\beta = 0.1$ to highly structured, interconnected macroscopic loops at $\beta = 0.5$. This contrast explicitly demonstrates the presence of the NHSE in the broader moir\'e continuum.}
    \label{fig:complex_spectrum}
\end{figure*}

In continuum moir\'e models, the effective physics is reliably captured by the low-energy states. Therefore, in this Appendix, we present a spectral analysis across an extended low-energy moir\'e continuum. This broader window allows us to explicitly visualize the NHSE in the dispersive bands, while simultaneously highlighting the remarkable robustness of the isolated flat band against non-Hermitian perturbations.

Figure~\ref{fig:complex_spectrum} illustrates this broader low-energy eigenvalue distribution for two representative non-Hermitian strengths, specifically weak ($\beta = 0.1$) and strong ($\beta = 0.5$) values of the nonreciprocity parameter, as shown in Figs.~\ref{fig:complex_spectrum}(a) and \ref{fig:complex_spectrum}(d), respectively. To systematically contrast the dispersionless and dispersive features, we enclose the target flat band near $E_0 \approx -17$~meV within a red rectangle and a highly dispersive moir\'e spectral region within a cyan rectangle. The magnified views of the isolated flat band (red rectangles) are presented in Figs.~\ref{fig:complex_spectrum}(b) and \ref{fig:complex_spectrum}(e). Although microscopic closed loops emerge at $\beta=0.5$ (preserving the nontrivial point-gap topology), the energy spread of the imaginary component of this flat band remains extremely constrained, confined strictly to an order of $10^{-7}$ to $10^{-5}$~meV. This clearly underscores the robust spectral pinning of the target flat band. 

In contrast, the dispersive low-energy moir\'e bands highlighted by the cyan rectangles exhibit a macroscopic evolution under the influence of non-Hermiticity, as shown in Figs.~\ref{fig:complex_spectrum}(c) and \ref{fig:complex_spectrum}(f). For $\beta=0.1$, the imaginary energy components remain relatively small (of the order of $10^{-3}$~meV), and the complex eigenvalues appear as a diffuse, scattered points. In this weakly non-Hermitian regime, the asymmetric hopping acts merely as a minor perturbation, and the bands do not enclose a significant area in the complex energy plane. However, at strong non-Hermiticity ($\beta=0.5$), the imaginary energy scale of these dispersive bands expands significantly by approximately anorder of magnitude (reaching up to $\sim 10^{-2}$~meV). More importantly, the formerly scattered eigenvalues systematically hybridize to form highly structured, interlocking complex-energy contours and multi-lobed patterns [see Fig.~\ref{fig:complex_spectrum}(f)]. According to the established non-Hermitian band theory, the formation of such braided, closed-loop structures under periodic boundary conditions is a standard signature of non-trivial point-gap topology \cite{PhysRevX.9.041015, RevModPhys.93.015005}. Under open boundary conditions, this topology dictates that a macroscopic number of these dispersive bulk states will exponentially localize at the boundaries of the system, confirming the robust presence of the NHSE \cite{PhysRevLett.121.086803, PhysRevLett.124.086801}. In summary, the formation of macroscopic spectral loops in the dispersive bands confirms the presence of the NHSE in the system with increase in the non-Hermitian parameter $\beta$. Moreover, the negligible imaginary part of the flat band demonstrates its inherent spectral stability against non-Hermiticity.

\section{Topological Phase Diagram in the $(\alpha, \theta)$ Plane}
\label{appendixc}

\begin{figure}[t]
    \centering
    \includegraphics[width=1\linewidth]{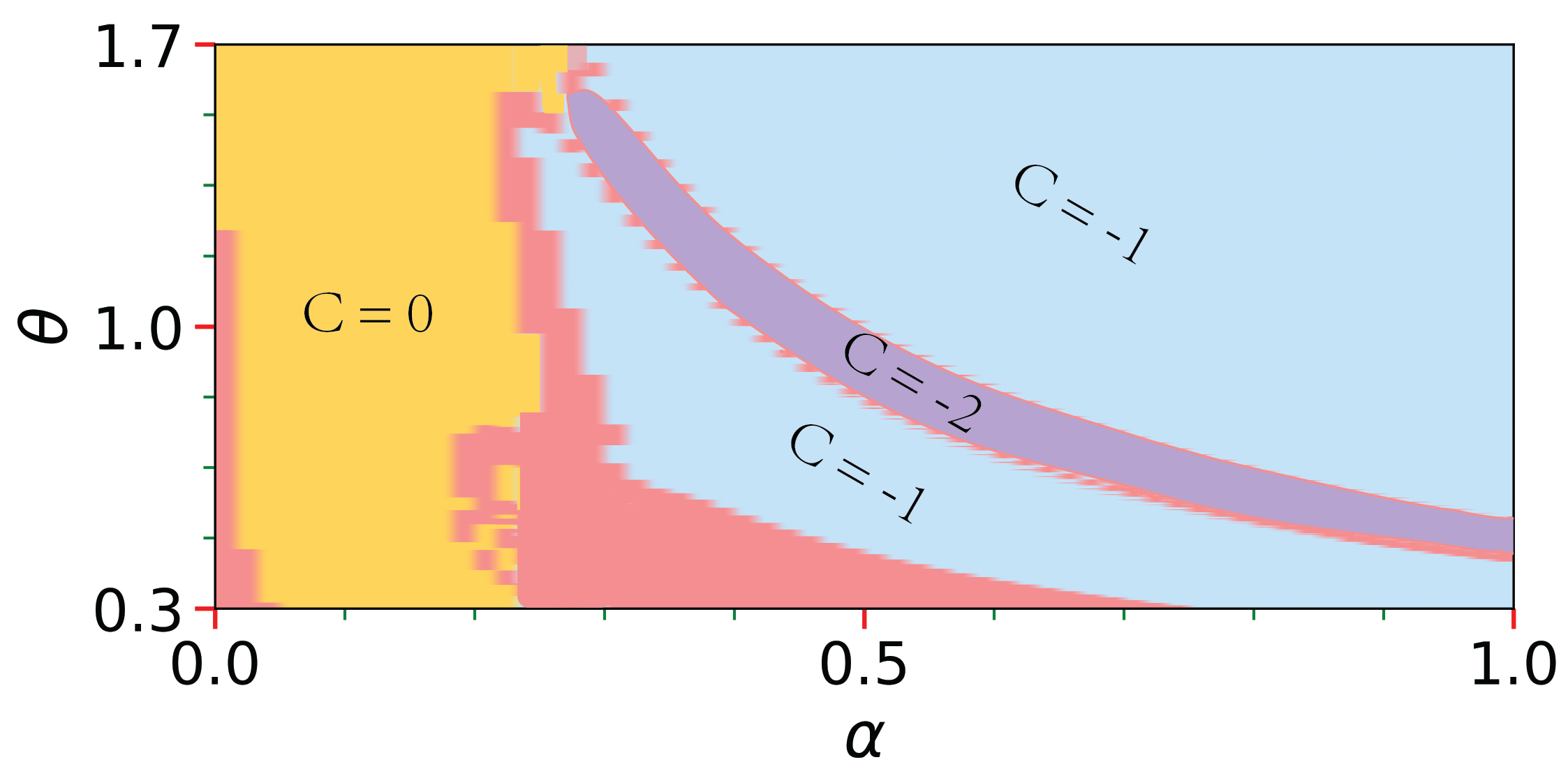}
   \caption{The biorthogonal Chern number $C$ is plotted as a function of the lattice parameter $\alpha$ and the twist angle $\theta$ for a fixed non-Hermiticity strength $\beta = 0.3$. Light-blue and purple regions represent topological phases with $C = -1$ and $C = -2$, respectively. Yellow areas denote the topologically trivial $C = 0$ phase. Light-red regions indicate gapless semimetallic states where the bulk gap $\Delta_R < 10^{-3}$~meV.}
    \label{fig:chern_alpha_theta}
\end{figure}

In the main text, we analyzed the topological phase transitions of the non-Hermitian twisted bilayer $\alpha-T_3$ lattice in the dice limit ($\alpha=1$) in Sec.~\ref{sec5}. To demonstrate the dependence of these topological states on the lattice geometry, we present the phase diagram of the biorthogonal Chern number in the $(\alpha, \theta)$ plane in Fig.~\ref{fig:chern_alpha_theta}. Here, the non-Hermitian parameter is fixed at $\beta = 0.3$. The light-blue and purple shaded regions represent the topological phases with biorthogonal Chern numbers $C = -1$ and $C = -2$, respectively. The light-red areas denote the gapless semimetallic regions ($\Delta_R < 10^{-3}$~meV) where the adjacent bands touch and the Chern number remains ill-defined. The yellow shaded region corresponds to the topologically trivial phase with $C = 0$. 

The phase diagram reveals a clear structural evolution of the topological states with respect to $\alpha$. Near the dice limit ($\alpha \to 1$), the system hosts well‑defined, gapped topological phases. Specifically, as the twist angle $\theta$ varies in this regime, the system transitions from a $C = -1$ phase to an intermediate $C = -2$ phase, and then back to a $C = -1$ phase. These distinct topological regions are physically separated by gap‑closing regions, across which the bulk energy gap vanishes. As $\alpha$ decreases toward intermediate values (approximately $0.3 < \alpha < 0.6$), the topological phase structure undergoes a significant shift. The $C = -2$ phase, which is prominent near the dice limit, begins to shrink in the $\theta$ direction. In this regime, as $\theta$ increases, the gapless boundaries (light‑red regions) surrounding the $C = -2$ phase converge. Eventually, these boundaries merge, leading to the annihilation of the $C = -2$ phase. Finally, for $\alpha < 0.2$, the broad gapless region terminates. The phase diagram shows that the band gap reopens across almost all twist angles, creating a large gapped region. In this regime, the system enters a topologically trivial phase with a Chern number of $C = 0$ (indicated by the yellow color). Physically, this occurs because the central C sublattice becomes nearly decoupled from the A and B sublattices at such low values of $\alpha$. Consequently, the specific flat band under consideration (near $E_0 \approx -17$~meV) reduces to a trivially flat and highly localized band, originating strictly from the completely isolated C sites.

\bibliographystyle{apsrev4-2}
\bibliography{References}

\end{document}